# Magnetic, magnetocaloric and transport properties of HoRuSi compound

Sachin B. Gupta,[1] K. G. Suresh,[1] and A. K. Nigam[2]

[1]Department of Physics, Indian Institute of Technology Bombay, Mumbai-400 076, India

[2]Tata Institute of Fundamental Research, Homi Bhabha Road, Mumbai-400 005, India

## Abstract

Magnetic, thermal, magnetocaloric and transport properties of polycrystalline HoRuSi have been studied. It was found that the compound shows magnetic transitions at $T_1$=18 K and $T_2$=8 K. Magnetization data reveal that the ferromagnetic ordering is dominant in this compound. Magnetocaloric effect has been estimated from M-H-T data and the $-\Delta S_M$ has been found to be 12.8 J/kg K for field of 50 kOe, which is comparable to some potential refrigerant materials in same temperature regime.

## 1. Introduction

Rare earth based intermetallics compounds with the composition RTX (R= rare earth, T= transition metal, and X is p block element) have been found to show very diverse and interesting magnetic and electrical properties. Depending on the constituent element various properties like superconductivity, heavy-fermion, large magnetocaloric effect, large magnetoresistance etc. have been observed in this family. The RRuSi compounds belong to this family and crystallize in $Co_2Si$ type orthorhombic crystal structure [1]. It has been observed that the compounds of RRuSi series show interesting properties and can be used for fundamental study as well as applications. Recently, we reported the magnetic and magnetocaloric properties of ErRuSi compound [2], which is ferromagnetic with a Curie temperature of 8 K. This compound has shown giant magnetocaloric effect (GMCE) around the ordering temperature, which was associated with the first order ferromagnetic to paramagnetic transition. Motivated by the results of ErRuSi, we have studied a few other compounds of this series in detail and the results on HoRuSi are presented in this paper.

## 2. Experimental details

Polycrystalline HoRuSi was synthesized by arc melting of constituent elements in a water cooled copper hearth under high purity argon atmosphere. The constituent elements, Er (99.9% at. purity), Ru and Si (99.99% at. purity) were melted by taking their stoichiometric proportion. The alloy button was melted four times to ensure homogeneity. As-cast sample was sealed in quartz tube in presence of vacuum ($10^{-6}$ torr.) and annealed for one week at 800 ℃, followed by furnace cooling to improve the homogeneity. The structural analysis was done by Rietveld refinement of the x-ray diffraction (XRD) patterns collected from X'PERT PRO at room temperature. The dc magnetization, heat capacity, and electrical resistivity measurements were performed in the PPMS (Quantum Design). The electrical resistivity was measured using standard four probe method, applying an excitation current of 150 mA. The ac susceptibility measurement has been carried out in MPMS, Squid VSM (Quantum Design).

## 2. Results and Discussion

The Rietveld refinement of the x-ray diffraction pattern shows no detectable impurity phases in the compound, confirming single phase nature of the compound. The lattice parameters, a=6.9145(2) Å, b=4.2550(2) Å, and c=7.1185(2) Å obtained from refinement are very close to reported values [1]. The xrd pattern along with the Rietveld refinement for HoRuSi compound is shown in Fig. 1.

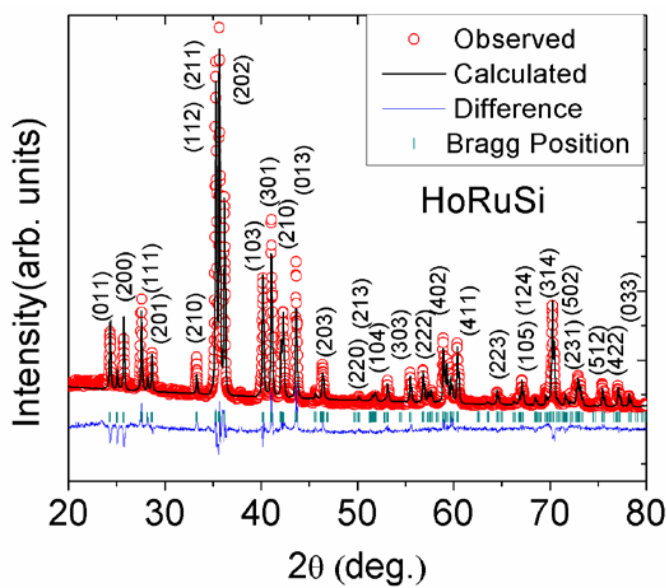

FIG. 1. Powder xrd pattern at room temperature along with the Rietveld refinement for HoRuSi compound. The bottom plot shows the difference between the observed and calculated patterns.

The temperature dependence of dc susceptibility (DCS) in zero-field-cooled (ZFC) and field-cooled (FC) conditions, obtained for the field of 500 Oe and 10 kOe is shown in Fig. 2 (left-hand scale). The DCS data shows only one clear anomaly at 18 K [Fig 2(a)]. One can see small thermomagnetic irreversibility at low temperature in DCS data measured at 500 Oe, arises due to pinning of domain walls. At high field this thermomagnetic irreversibility is removed. The inverse susceptibility has been plotted in Fig. 2(a, right-hand scale) and fitted with Curie-Weiss law in paramagnetic region. The observed effective magnetic moment ($\mu_{eff.}$) and paramagnetic Curie temperature ($\theta_p$) estimated from Curie-Weiss fit to inverse susceptibility data have been found to be 10.8 $\mu_B$/Ho$^{3+}$ and 13 K. The observed value of $\mu_{eff.}$ is close to free Ho$^{3+}$ moment value ($\mu_{th.}$=10.6 $\mu_B$), which indicates that there is no moment on the Ru atom, if exists, is very small. The positive value of $\theta_p$ suggests that ferromagnetic interaction is predominant in this compound.

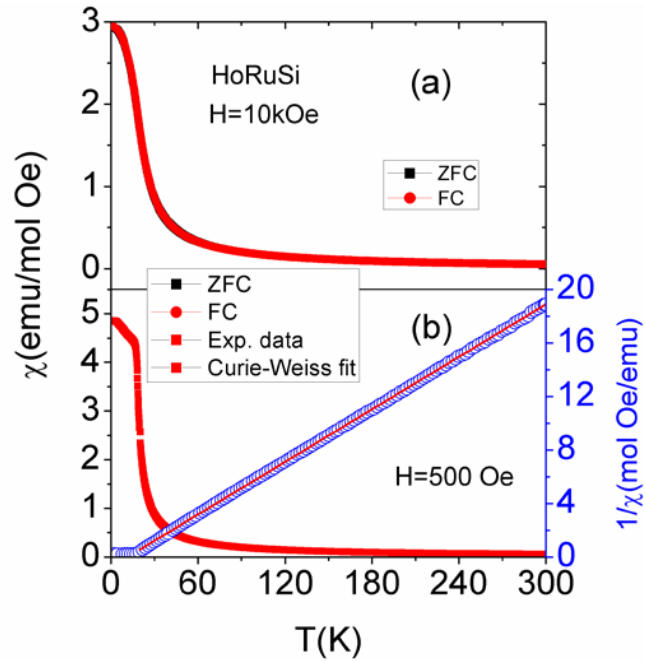

FIG. 2. Temperature dependence of dc magnetic susceptibility for HoRuSi at 500 Oe and 10 kOe. The right hand scale in (a) shows the Curie-Weiss fit to inverse susceptibility data.

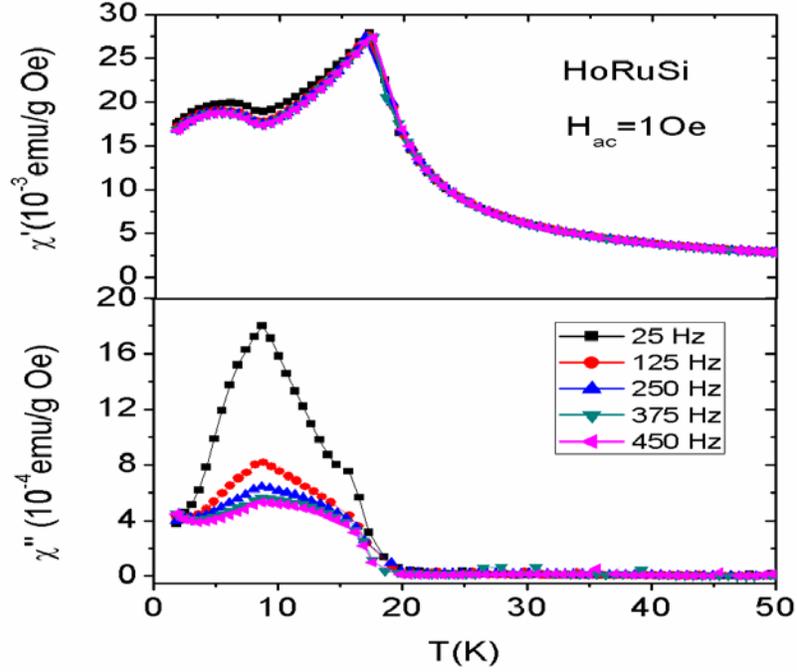

FIG. 3. Temperature dependence of ac susceptibility carried out at various frequencies (f=25-625 Hz) and constant ac field ($H_{ac}$=1 Oe).

To know the detailed magnetic phase transition, we have carried out the ac susceptibility measurement at different frequencies (f=25-625 Hz) and constant ac field ($H_{ac}$= 1 Oe). The ac susceptibility (ACS) measurement is the best method for the study of magnetic dynamics and precise determination of magnetic phase transition in the magnetic system. The ac measurements under ZFC condition with small amplitude of oscillating field $H_{ac}$ and frequencies f do not disturb a magnetic system significantly, thus, providing precise determination of magnetic phase transition temperatures [3]. This method has also been used to study the non linear behavior of magnetic systems [3]. The temperature dependence of ZFC in-phase $\chi'(f,T)$ and out-of-phase susceptibility $\chi''(f,T)$ recorded simultaneously is shown in Fig. 3. There are two peaks, one near 17.8 K ($T_1$) (corresponding to DCS data) and other at 8 K ($T_2$) in both $\chi'(f,T)$ and $\chi''(f,T)$ of ACS. The presence of peaks in $\chi''(f,T)$ near $T_1$ and $T_2$ indicates the ferromagnetic ordering in this compound. Additionally, it can also be observed from Fig. 3 that there is influence of frequency on the real part ($\chi'$) as well as imaginary part ($\chi''$) of ACS, which reflects different energy losses in magnetically ordered system depending upon f [4]. Such losses are the characteristics for the magnetic systems having a net magnetic moment (e.g. ferromagnetic, ferromagnetic, or canted systems, spin glasses) [5-7]. Hence the ac susceptibility investigation made on HoRuSi reveals order-order transition of the compound.

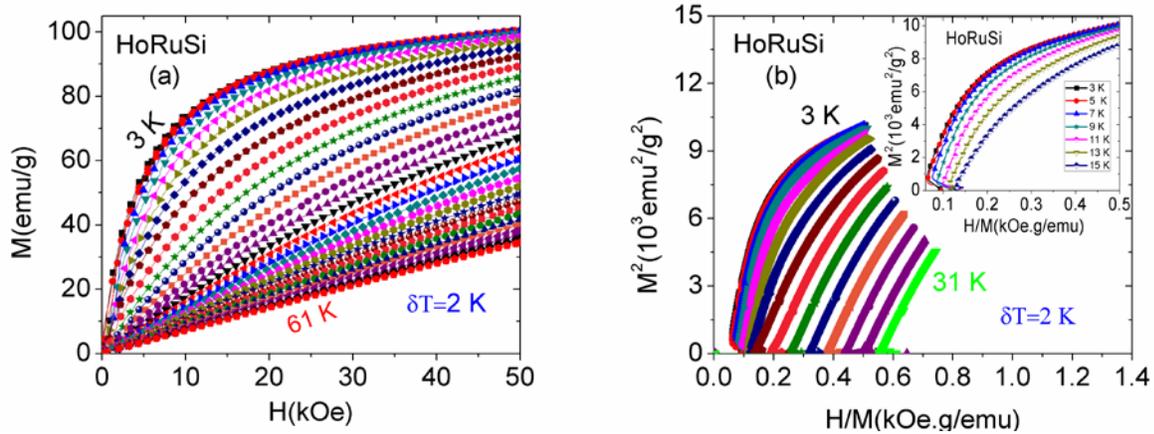

FIG. 4. (a) The field dependence of magnetization at different temperatures recorded for field up to 50 kOe. (b) The Arrott plots at selected temperatures for HoRuSi. Inset in (b) shows an expanded scale view of Arrott plots for HoRuSi.

Fig 4(a) displays the magnetization isotherms in the temperature range 3-61 K with a temperature interval of 2 K for field up to 50 kOe. It is clear from the M (H) curves that the magnetization decreases with increase in temperature and become linear above ordering temperature. This indicates that ferromagnetic ordering is dominant in this compound. The magnetization isotherms give an indication of saturation of moments at higher field. The saturation moment at 3K and 90 kOe is found to be 5.75 $\mu_B$, which is smaller than saturation moment for single $Ho^{3+}$ ion (gJ=10 $\mu_B$). The reason behind the smaller value of saturation moment is not clear yet. To get a clear idea about decrease in saturation moment, neutron diffraction measurement is required, which is in progress. It can be noted from Fig 4(a) that there is negligible thermal and field hysteresis in this compound, which is advantageous for magnetic refrigeration purpose. A large reversible MCE is expected near the ordering temperature where the magnetization changes rapidly with varying temperature. To know the order of magnetic phase transition, the magnetization isotherms were plotted in the form of Arrott plots and are shown in Fig 4(b). According to Banerjee criterion,[8] the negative slope of the Arrott plot shows the first order magnetic transition while the positive slope indicates the second order transition. The closer view of Arrott plots [see the inset of Fig 4(b)] shows a negative slope of Arrott plots at low temperatures and fields indicates the first order magnetic phase transition at low temperatures and fields.

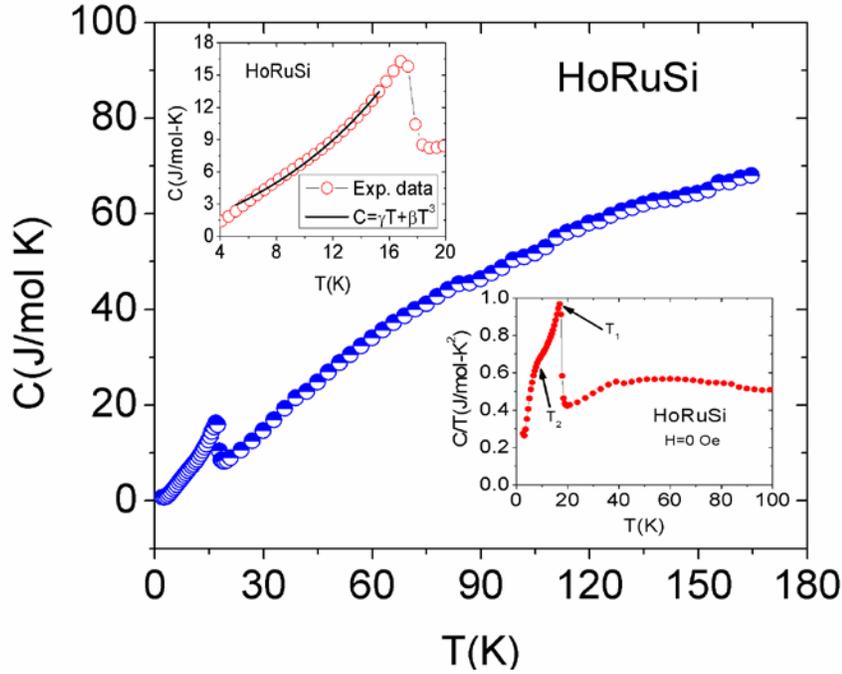

FIG. 5. Temperature dependence of heat capacity measured in zero field. Inset show the heat capacity data on an expanded plot.

To understand the system in detail, the heat capacity measurement was done at zero field and in the temperature range of 2-100 K. The heat capacity in presence of field could not do because of the pulling out of the sample in presence of field from heat capacity puck. The heat capacity data shows a sharp peak at 18 K (see Fig. 5) and a hump near $T_2$ (see the lower inset of Fig 5). The sharp peak at 18 K reveals that the compound shows the first order phase transition as also suggested by Arrott plots. A small hump at $T_2$ may arise due to the ferro to ferromagnetic transition in this compound. The upper inset in Fig 5 shows the fit of equation, $C=\gamma T+ \beta T^3$ to C-T data at low temperatures, which yields $\gamma$=533 mJ/mol-$K^2$ and $\beta$= 1.49 mJ/mol-$K^4$. The value of Debye temperature, $\theta_D$ has been estimated from formula $\sqrt[3]{1944/\beta}$ and is 109 K.

The magnetocaloric effect in this compound has been evaluated from magnetization data using Maxwell's relation given as

$$\Delta S_M = S(T,H) - S(T,0) = \int_0^H \left[\frac{\partial M}{\partial T}\right]_H dH \qquad (1)$$

Here, M is magnetization and H is the applied field. Using the magnetization data at different temperatures and fields, the change in magnetic entropy can be calculated by the simplified formula [9]

$$\Delta S_M = \sum_i \frac{M_{i+1} - M_i}{T_{i+1} - T_i} \Delta H_i \qquad (2)$$

Where, $M_{i+1}$ and $M_i$ are the magnetization values at temperature $T_{i+1}$ and $T_i$ in a field $H_i$ respectively. The change in magnetic entropy with temperature at different fields is shown in Fig 6. The MCE data shows a large peak near $T_1$, while the careful observation also shows a second peak around $T_2$. At this temperature the change in $\Delta S_M$ is very sharp, it decreases sharply with temperature. One can see that the maximum of - $\Delta S_M$ lies near $T_1$.

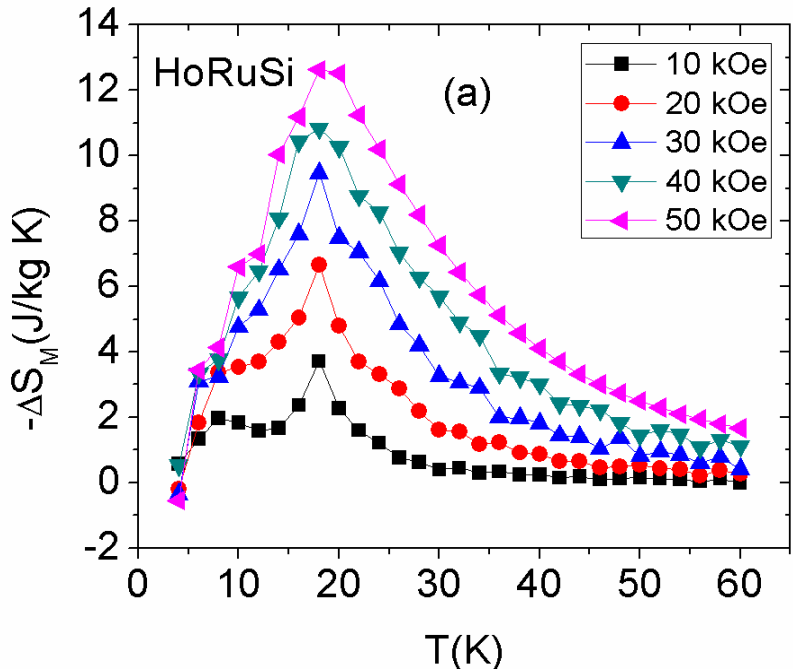

FIG. 6. (a) Temperature dependence of $\Delta S_M$ for HoRuSi. .

The peak value of $-\Delta S_M$ around $T_1$ is 12.8 J/kg K for the field change of 50 kOe. It is worth to note the MCE of HoRuSi compound is comparable to Ho compounds of other series in RTX family; such as HoRhSn (13.2 J/kg K) [10], HoRhGe (11.1 J/kg K) [11], and HoRhSi (11.7 J/kg K) [12] subjected to same field changes and temperature below 20 K. Here, one can also note that replacing Rh atoms by Ru resulted in small increase in MCE. The value of MCE is also comparable to some proposed potential magnetic refrigerant materials such as $Gd_2PdSi_3$ (8 J/kg K at 40 kOe) [13], $DyCO_3B_3$ (12.6 J/kg K) [14], $HoMnO_3$ (13.1 J/kg K at 70 kOe) [15], $Ho_3Co$ (14.5 J/kg K) [16], working in same temperature regime. The other important parameter for magnetic refrigeration is the refrigerant capacity (RC) or relative cooling power (RCP), which is the measure of heat transfer from cold reservoir to hot reservoir in an ideal refrigeration cycle. The RC is defined as the product of $\Delta S_M^{max}$ and the full width at half maximum of $\Delta S_M$ vs. T plot [3].

$$RC = -\Delta S_M^{max} \delta T_{FWHM} \qquad (3)$$

The RC value of HoRuSi for field change of 50 kOe is found to be 296 J/kg and is comparable to good refrigerant materials working in same temperature regime. Thus, the comparison of MCE parameters with other potential materials suggests that the compound is promising for magnetic refrigeration at low temperatures.

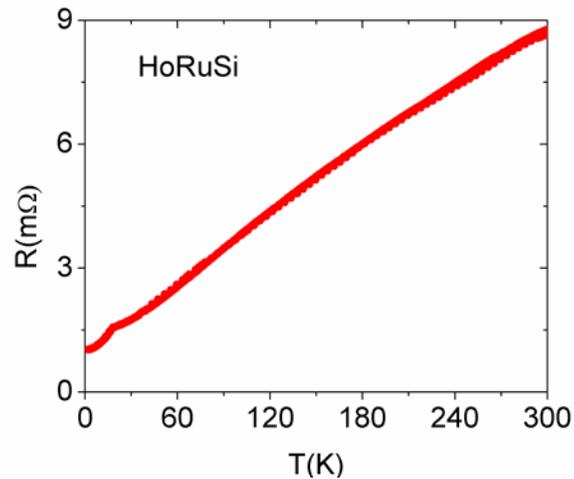

FIG. 7. Temperature dependence of electrical resistance for HoRuSi.

Fig. 7 shows the temperature dependence of the resistance for title compound. The compound shows slope change at 18 K. there is no anomaly associated with $T_2$ in R data.

## Conclusions

HoRuSi shows double magnetic transitions and seems to be dominantly ferromagnetic as suggested by magnetic, caloric, magnetocaloric, and resistivity studies. Large value of MCE was observed in this compound for the field change of 50 kOe, which is comparable to many potential magnetic refrigerants and Ho compounds of RTX family. The compound shows first order magnetic phase transition at low temperature and fields.

## Acknowledgments

S.B.G. thanks CSIR, Govt. of India for granting senior research fellowship. S. B. G. is grateful to Dr. A. Das for fruitful discussion with him. Authors would like to thank D. Buddhikot for his helping hands in resistivity measurements.

## References

[1] A. V. Morozkin, Y. D. Seropegin, I. A. Sviridov, and I. G. Riabinkin, J. Alloys Compd. **282**, L4 (1999).

[2] S. B. Gupta and K. G. Suresh, Appl. Phys. Lett. **102**, 022408 (2013).

[3] K. Łątka, R. Kmieć, A. W. Pacyna, R. Mishra, and R. Pöttgen, Solid State Sci. **3**, 545 (2001).

[4] K. Łątka, R. Kmieć, R. Kruk, A. W. Pacyna, T. Fickenscher, R. D. Hoffmann, and R. Pöttgen, J. Solid State Chem. **178**, 2077 (2005).

[5] T. Sato, T. Ando, T. Watanabe, S. Itoh, Y. Endoh, and M. Furusaka, Phys. Rev. B **48,** 6074 (1993).

[6] K. Binder and A. P. Young, Rev. Mod. Phys. **58,** 801 (1986).

[7] H. Maletta, W. Zinn, Spin glasses, in: K. A. Gschneidner, Jr., L. Eyring (Eds.), Handbook on the Physics and Chemistry of Rare Earths, North-Holland, Amsterdam **12**, 213 (1989).

[8] S. K. Banerjee, Phys. Lett. **12**, 16 (1964).

[9] H. Zhang, Z. Y. Xu, X. Q. Zheng, J. Shen, F. X. Hu, J. R. Sun, and B. G. Shen, J. Appl. Phys. **109,** 123926 (2011).

[10] S. B. Gupta, K. G. Suresh, and A. K. Nigam, J. Appl. Phys. **112**, 103909 (2012).

[11] S. B. Gupta, K. G. Suresh, and A. K. Nigam, communicated.

[12] S. B. Gupta, K. G. Suresh, and A. K. Nigam, communicated.

[13] E. V. Sampathkumaran, I. Das, R. Rawat, and S. Majumdar, Appl. Phys. Lett. **77**, 418 (2000).

[14] L. Li, H. Igawa, K. Nishimura, and D. Huo, J. Appl. Phys. **109**, 083901 (2011).

[15] A. Midya, P. Mandal, S. Das, S. Banerjee, L. S. S. Chandra, V. Ganesan, and S. R. Barman, Appl. Phys. Lett. **96**, 142514 (2010).

[16] J. Shen, and J. F. Wu, J. Appl. Phys. **109**, 07A931 (2011).